# Effect of Strain Relaxation and the Burstein-Moss Energy Shift on the Optical Properties of InN Films Grown in the Self-Seeded Catalytic Process

*Kishore K. Madapu,[*] and Sandip Dhara[*]*

Surface and Nanoscience Division, Indira Gandhi Centre for Atomic Research, Kalpakkam–603 102, India.

[*] Electronic Mail: madupu@igcar.gov.in; dhara@igcar.gov.in

*Abstract*

For the first time, high optical quality InN films were grown on sapphire substrate using atmospheric chemical vapour deposition technique in the temperature range of 560–650 ºC. Self-catalytic approach was adopted to overcome the nucleation barrier for depositing InN films. In this process, seeding of the nucleation sites and subsequent growth was performed in the presence of reactive $NH_3$. We investigated the simultaneous effect of strain and Burstein-Moss (BM) energy shift on optical properties of InN films using Raman and photoluminescence spectroscopy. Existence of compressive strain in all films is revealed by Raman spectroscopic analysis and is found to relax with increasing growth temperature. The asymmetric broadening of the $A_1$(LO) phonon mode is observed with the onset of plasmon–phonon interaction for films grown at 620 ºC. Large blue shift of the band gap of InN (1.2 eV) is observed as a collective result of compressive strain in films as well as BM shift. Carrier density is calculated using the BM shift in the photoluminescence spectra.

Finally, blue shift in band edge emission is observed further because of the presence of compressive strain in the films along with the BM effect.

# 1. Introduction

InN, one of the members of III-nitride family, has been a less studied material as compared to other nitrides, namely GaN and AlN, because of its difficulties in growing the high quality crystalline material.[1] However, research interest on InN is not tarnished because of its superior electronic properties such as high carrier mobility and low effective electron mass (0.07 $m_e$).[2] After a decade long debate, band gap of InN has recently been settled in the near infra-red region (0.75 eV) from 1.9 eV giving rise its applicability to near infrared and terahertz region.[3,4] However, studies related to effect of strain and carrier density on optical emission properties are inadequate. As of now, most of the InN films are deposited by controlled growth techniques such as molecular beam epitaxial (MBE),[5-7] and metal organic vapor phase epitaxy (MOVPE).[8] Moreover, S. Zhao *et. al.*[9] was also able to synthesis *p*-type InN nanowires by improved MBE technique. The deposition and decomposition temperatures of the InN phase occur in the same range (500-650 $^o$C) and it is comparatively easy to control such condition in the vapor phase epitaxial technique. As a result of this, one fails to found any reports on optical grade InN films in the atmospheric chemical vapour deposition (APCVD) technique even though it is a commercially viable technique for large scale synthesis. In earlier reports, InN was synthesized in CVD technique with the conversion of $In_2O_3$ by nitridation using the $NH_3$ and sublimation of $NH_4Cl$ with In.[10-12] In these reports, optical emission properties were not reported, suggesting low optical quality in these samples.

As in the case of other III-nitrides, InN also suffers from lack of native substrate. Usually, InN is grown on sapphire substrates with buffer layer of GaN or AlN.[8] Generally, foreign substrates are attributed with lattice mismatch because of difference in the lattice constants and the thermal expansion coefficients. However, the developed strain may relax by forming the misfit threading dislocations in the films,[13] leading to the degradation of electronic and optical properties of the material. Recently, researchers reported growth of InN on the In$_2$O$_3$(111)[14] and ZnO (000-1)[15] as alternate substrates. In addition, thick microcrystals of InN were grown on *c*-plane sapphire with Mo mask to reduce the threading dislocations.[16] Nevertheless, nanostructures, especially nanowires, alleviate the strain and dislocation effects because of their effective strain relaxation in the presence of high surface to volume ratio.[17,18] In the present case, in addition to the difference in the thermal expansion coefficient, there is a huge lattice mismatch between the InN and sapphire substrates (27%). As a result, the residual strain is expected in the grown films. Thermal expansion mismatch induced strain is developed during the cooling of the sample from the growth temperature to the room temperature. Therefore, strain developed by thermal expansion mismatch can be expressed as $\varepsilon = (\alpha_{substrate} - \alpha_{film}) \cdot \Delta T$ where $\alpha$ is the thermal expansion coefficient and $\Delta T$ is the difference in growth temperature to room temperature. Usually, lattice and thermal expansion mismatch induced strain in GaN films, grown on sapphire substrate, is biaxial and compressive in nature. In addition to it, point or extended defects contribute as hydrostatic strain in the films.[19,20] Hydrostatic strain is either compressive or tensile depending on the size of the atoms involved in the defect formation.[12] However, strain effect on the optical properties of GaN is well established. A blue shift of the emission peak is initiated by the compressive strain and correspondingly a red shift occurs due to the presence of tensile strain as compared to strain free samples.[21-23] Reports on strain effect on optical properties of InN, however are scarce because of lack of high quality materials. In this context, Raman spectroscopy is established as a simple and effective technique to evaluate strain in the films by measuring the shift in the phonon frequencies. In addition to this, especially in case of III-nitrides, Raman spectroscopy is useful to study the coupling of charge carriers with LO phonon mode through Frölich interaction.[6, 24]

In addition to strain, blue shift of band gap energy because of the fact that Burstein-Moss (BM) phenomenon plays a crucial role in the optical properties of InN because of the high amount of carrier density.[25] Moreover, high surface electron accumulation of InN films has its own effect on the optical properties.[26-28] The shift due to the BM effect is observed due to the occupation of the higher energy levels in the conduction band from where the electron transition occurs instead of the conduction band minimum. On account to the BM effect, optical band gap is virtually shifted to high energies because of the high carrier density related band filling. Thus, studies are required to establish the simultaneous effect of the strain and the carrier density on optical properties of InN.

In the present work, successful deposition of InN is reported using the self-catalytic approach to overcome the difficulties in the synthesis of InN. Effects of growth temperature, in the range of 560-650 °C, on strain and carrier density of the InN films are studied by analysing the phonon modes using the Raman spectroscopy. Subsequently, optical emission properties depending on the strain and carrier density are studied by photoluminescence (PL) spectroscopy at 80 K. In addition, temperature dependent PL study is also carried for these samples to shed light on the electron-hole recombination process.

## 2. Experimental

InN films were deposited using the APCVD technique in the temperature range of 560 – 650 °C. Metallic In (99.999%) shots and ultra high pure $NH_3$ (99.9999%) were used as the source and reactive gases, respectively. Growth was carried in a custom designed CVD set up with a horizontal tube furnace.[29-30] The *c*-axis oriented (0001) sapphire ($Al_2O_3$) substrates of dimension $0.5 \times 0.5$ cm$^2$ were used without any buffer layer. Alumina boat (99.95%) was used for holding the source and substrates. In the present case, growth and decomposition temperatures of InN were in the same range. In order to have effective deposition, special arrangements of source material and substrate were adopted such that the substrate was surrounded by four In shots at its corner with the 0.5 cm separation. Subsequently, ceramic boat with source and substrate was transferred to 1 inch quartz tube for the deposition. The thermocouple was placed very close to quartz tube for accurate measurement of temperature in the chamber. In order to carry out the growth, two different schemes were followed to reach the growth temperature: in one case 1) the temperature was directly increased to the growth temperature with the ramp rate of 20 $^{o}C \cdot min^{-1}$ and in another case 2) the temperature was increased to 500 $^{o}C$ first to stay for 30 minutes before reaching the growth temperature with same ramp rate, 20 $^{o}C \cdot min^{-1}$. The reaction chamber was maintained at the base pressure of $10^{-3}$ mbar with a rotary pump evacuation until the growth temperature was reached. Once the growth temperature was stabilized, the reacting gas of $NH_3$ was started flowing with a flow rate of 100 sccm. Growth was carried for 2 hr. After the growth, furnace was cooled to room temperature while maintaining the $NH_3$ atmosphere up to 400 $^{o}C$. Successful deposition of InN films was observed in the second scheme of temperature ramping. Complete growth mechanism will be discussed later in the results and discussion section.

Topography of the grown films was found by tuning fork based atomic force microscopy (AFM, Mutiview 4000, Nanonics) with a tip size of 20 nm. Lattice vibrational properties were studied using Raman spectrometer (inVia, Renishaw) in the backscattering configuration with excitation of 514.5 nm laser. Scattered light was collected in the thermoelectric cooled CCD detector after dispersion by an 1800 $gr \cdot mm^{-1}$ grating. Optical properties of the samples were studied using temperature dependent PL spectroscopy with the excitation of a 785 nm diode laser, using 600 $gr \cdot mm^{-1}$ grating and a single channel InGaAs photo-detector in the similar backscattering configuration.

## 3. Results and Discussion

### 3.1 AFM Topography

Typical topography of nanocrystalline films grown in the range of 560-650 °C using the second scheme are shown in the figures 1a-d. It is reported that the III-nitride thin films, grown in the MOCVD technique on the *c*-axis oriented sapphire substrate, possess the mosaic structure.[31] In the present case also, topographic images illustrates mosaic nature in the grown InN films. Thicknesses of the samples, as measured using AFM, are in the range of 0.5 - 0.7 μm. Typical root mean square (*rms*) roughness value of 17 nm (Fig. 1a) is observed for 560 °C grown sample. In case of 580 °C (Fig. 1b) and 600 °C (Fig. 1c) grown samples, *rms* roughness were found to be 44 and 43 nm, respectively. High temperature grown sample, at 650 °C, shows the larger grain size (Fig. 1d) with the *rms* roughness of 35 nm.

**3.2 Growth mechanism**

As mentioned earlier deposition of high quality InN material is difficult in the CVD technique because of the fact that the formation and decomposition temperatures of InN fall in the same range (550-650 °C). Nucleation of InN phase at these temperatures is very difficult because of the occurrence of simultaneous reversible reaction (InN→In(*s*) + $N_2$(*g*)). Figure 2a shows growth scheme carried using first kind of the ramp rate. In this case, InN deposition was not observed because of nucleation barrier hinders the subsequent growth. In other words, simultaneous introduction of reactants and nucleation of InN is impossible in the same range of temperature. To overcome the nucleation barrier, the nucleation sites are seeded using In before introducing the reactive $NH_3$. As shown in the figure 2b, there are four processes, represented by steps A to D, to achieve the InN deposition. Process A involves the ramping of the temperature to a stable temperature 500 °C. Step B involves the seeding of the nucleation sites using the In for 30 minutes. Process C involves the introduction of reactive $NH_3$, which is decomposed into the atomic N and $H_2$. Subsequently, nucleation of InN is initiated at pre-deposited In nucleation sites. As the time goes, the growth of InN is observed with grain growth. This growth mechanism typically follows the island (Volmer-Weber) growth mechanism. Subsequently coalescence of island is observed with the possible occurrence of Ostwald ripening process. As a result of this, AFM images show the nanocrysatlline (mosaic) nature of the grown films. However, deposition of InN is not observed above the growth temperature 650 °C even though nucleation sites are seeded, as dissociation of InN takes place at high temperature. On the contrary, poor crystalinity of the grown material is expected in case of growth temperature below 560 °C, as the compositional homogeneity may not reach at this temperature of early nucleation stage. To substantiate our growth mechanism, we have taken out the substrate after the process B to study its topography. Figure 3a shows the morphology of the sapphire substrates prior to the deposition. However, figure 3b show the sapphire substrate after the process B where tiny particles are clearly observed uniformly over the substrates as a result of the seeding of In nucleation sites.

**3.3 Raman spectroscopic study**

Hexagonal wurtzite InN belongs to the space group of $p6_3mc(C_{6v}^4)$ with four atoms in the primitive unit cell. In this structure each atom occupies the $C_{3v}$ sites. At zone centre ($q$=0), group theory predicted irreducible representation for this structure is given by $\Gamma$ = $A_1+E_1+2E_2+2B_1$.[32,33] Among these, both $A_1$ and $E_1$ phonon modes are infra-red (IR) and Raman active. In contrast to this, $E_2$ phonon modes of $E_2$(high) and $E_2$(low) are exclusively Raman active. However, $B_1$ phonon mode is neither Raman nor IR active and is termed as forbidden mode. The $A_1$ and $E_1$ phonon modes are polar in nature and these vibrations polarize the unit cell leading to further split in to longitudinal optical ($A_1$(LO) and $E_1$(LO)) and transverse optical ($A_1$(TO) and $E_1$(TO)) modes. Consequently, the wurtzite phase of InN consists of six Raman active phonon modes namely, $2A_1$, $2E_1$, and $2E_2$ at zone centre.

Figure 4a shows the typical Raman spectra of InN films grown at different temperatures. These spectra are recorded at room temperature with the laser power of 0.3 mW to avoid the laser heating effect. Raman spectra shown in figure 4a are without any correction for the background. In the present study, the Raman spectra were collected in the back scattering configuration of $Z(X-)\bar{Z}$. Here $Z$ is the direction of propagation of incident light which is parallel to the *c*-axis of the *wurtzite* structure of InN and $X$ represents the direction of polarization of the electric field. For the backscattering configuration of $Z(X-)\bar{Z}$ and $Z(Y-)\bar{Z}$, the allowed Raman modes are $E_2$ and $A_1$(LO) only. The Raman spectra shows distinct peaks in the range of 84–85 cm$^{-1}$, 490–493 cm$^{-1}$ and 589−593 cm$^{-1}$ which correspond to the $E_2$(low), $E_2$(high) and $A_1$(LO) phonon modes, respectively. For the case of 560 °C grown film, Raman modes corresponding to sapphire substrate are also observed in the spectrum which are indicated by star (*) marks. Origin of a tinny peak centred at 444 cm$^{-1}$, which can be assigned either as $A_1$(TO) mode or a longitudinal optical phonon plasmon coupling mode (LOPCM),[7] is discussed later. However, in the backscattering configuration $A_1$(TO) mode is not allowed.

The $E_2$(high) phonon mode is a best measure of the strain and crystalline quality of the films because of the reason that it is not affected by the carrier density in the material for its non-polar nature.[13, 34] Shift in the $E_2$(high) phonon mode frequency depends on the nature of strain, either being compressive or tensile, present in the films. In other words, as a result of compressive strain a blue shift in the $E_2$(high) mode is observed. Moreover, broadened full width half maxima (FWHM) of the $E_2$(high) phonon mode reveals the possible presence of the native defects.[19] Figure 4b shows the spectral region of the $E_2$(high) phonon mode of InN films grown at different temperatures along with the Lorentzian fits. The accurate spectral frequency and FWHM corresponding to the $E_2$(high) phonon mode were determined by fitting with the Lorentzian function. Frequency of the $E_2$(high) phonon mode is blue shifted in all films with respect to the strain free films grown on sapphire in the MBE technique as reported by Davydov et al.,[35] where $E_2$(high) mode is observed at 488 cm$^{-1}$. Demangeot et al.,[34] also reported similar blue shift in the InN micro-dots grown on sapphire substrate with GaN buffer layer. Blue shift of the $E_2$(high) phonon mode with respect strain free values reveals the presence of compressive strain in the films. Here, the compressive strain in the films is attributed to large mismatches of lattice and thermal expansion coefficients of the substrate and films (lattice parameters: $a_{InN}$ = 3.54 Å, $c_{InN}$ = 5.70 Å, $a_{sapphire}$ = 4.758 Å, $c_{sapphire}$ = 12.991 Å; thermal expansion coefficient in $a$ plane: $a_\perp$ = 4×10$^{-6}$ and 7.5×10$^{-6}$ K$^{-1}$ for InN and sapphire, respectively). From the fitting of $E_2$(high) phonon mode, red shift in the frequency is observed for increasing growth temperature (Fig. 5a, left side scale) in the present report. In addition to this, FWHM of the $E_2$(high) phonon mode is increased with increasing growth temperatures (Fig. 5a, right side scale). Film grown at 560 °C has shown the opposite trend for the variation of the FWHM. It may be because of the fact that low crystalline quality material is produced at relatively low growth temperature. Finally, $E_2$(high) mode frequency is red shifted by 1.82 cm$^{-1}$ when growth temperature increases from 560 to 650 °C.

Thermal and lattice mismatch induced strain in the InN films is usually biaxial in nature.[22] In case of biaxial strain, according to the symmetry considerations, there are only three non-vanishing strain component which are given by

$$\varepsilon_{xx} = \varepsilon_{yy} = \left(\frac{a-a_0}{a_0}\right) \quad \text{............................................................ (1)}$$

$$\varepsilon_{zz} = \left(\frac{c-c_0}{c_0}\right) = -\frac{C_{13}}{C_{33}}(\varepsilon_{xx} + \varepsilon_{yy}) \quad \text{............................................ (2)}$$

$$\varepsilon_{zz} = -2\frac{C_{13}}{C_{33}}(\varepsilon_{xx}) \quad \text{...............................................................} \quad (3)$$

where $C_{ij}$ are the stiffness coefficients. The strain induced phonon shift within the limit of Hooke's law is expressed by,

$$\Delta\omega = \omega - \omega_0 = 2a_{E_2}\varepsilon_{xx} + b_{E_2}\varepsilon_{zz} = 2\left[a_{E_2} - b_{E_2}\frac{C_{13}}{C_{33}}\right](\varepsilon_{xx}) \quad \text{.....................} \quad (4)$$

where coefficients $a_{E_2}$ and $b_{E_2}$ in the Eqn. (4) are phonon deformation potentials. However, deformation potential of InN is reported with values very much scattered in the literature.[5,36] In present case, the strain was calculated by taking the deformation potentials as reported by Kim et al.,[37] i.e., $a_{E_2} = -998$ cm$^{-1}$, $b_{E_2} = -635$ cm$^{-1}$ and stiffness coefficients of $C_{13}$=124, and $C_{33}$=200. Figure 5b shows the variation of strain with respect to the growth temperature. Strain is observed to decrease with increasing growth temperature. In other words, the compressive strain, produced by thermal and lattice mismatch, is relaxed at the higher growth temperatures. Moreover, the FWHM of the films is also found to increase with increasing growth temperature except for films grown at 560 °C, as shown in Fig. 5a (right side scale). This observation unambiguously indicates that defect density is increased with the increased growth temperature. Strain produced from the point defects is hydrostatic in nature, which is either compressive or tensile depending on the involved atom sizes. In the present case, partial decomposition of InN is more likely to happen at high growth temperatures.[29] As a result of this, formation of defect may be expected for the films grown at relatively high temperature leading to the formation of supersaturation of In and In$_N$ antisite defects with higher concentration.[13] Subsequently, as a result of structural imperfections, strain relaxation takes place which is corroborated with the red shift of $E_2$(high) phonon mode along with the broadening of FWHM with increasing growth temperature. Recently, we reported evolution of InN quantum dots to nanorods using $In_2O_3$ as precursor in APCVD technique,[29] where presence of large amount of O is observed in the samples grown at high temperatures. The dissociation of InN at high temperature and subsequent reconversion to $In_2O_3$ was reported in the presence of O, originating possibly from the precursor. In the present study, however presence of O impurity is ruled out because of the absence of any oxide phase in the precursor material and the utilization of ultra high pure $NH_3$. As discussed earlier, increase in the growth temperature is associated with partial dissociation of InN. As a result of the dissociation, number of N vacancies ($V_N$) are increased, which results in the creation of point defects (imperfections) in the grown films. This assumption coupled the occurrence of $V_N$ is corroborated from the observed increase in the carrier density for high temperature grown samples.[2] The increase in carrier density is supported by Raman spectroscopic analysis of $A_1$(LO) phonon mode of films grown at high temperature in the subsequent section. In the present case, the hydrostatic strain originated in the presence of $V_N$ is tensile in nature. Consequently, a compressive strain induced by the foreign substrate is relaxed by the tensile strain produced by the $V_N$. As a net result, $E_2$(high) phonon mode is found to be red shifted along with an increase in the FWHM for the films grown at high temperatures.

In polar semiconductors, LO phonon modes are coupled with charge carriers through the macroscopic electric fields produced by these vibrations. Thus, we claim to observe the LOPCMs such as the $\omega_\pm$ represented by $L^+$ and $L^-$, respectively. Moreover, behaviour of LO phonon modes in case of InN are well studied by several groups for different carrier concentrations.[6,7,32,38,39] However, it is still under debate over the years because of contradicting results in the literature. The frequencies of $L^-$ and $L^+$ fall in the range of 400-450 and ~885 $cm^{-1}$, respectively. However, these frequencies are strongly dictated by the carrier density of the system. Nevertheless, high frequency coupled mode ($L^+$) is seldom observed experimentally in the Raman spectra. In the present case, the peak observed at 445 cm$^{-1}$ is assigned as the $A_1$(TO) phonon mode looking into its insensitiveness to carrier concentration of the samples.[7] Appearance of $A_1$(TO) mode, which is forbidden in the backscattering configuration, is attributed to the breakdown of polarization selection rules due to the misorientation of crystal with respect to laser direction. However, line shape of un-screened LO phonon mode is observed to depend on the charge carrier density.[40-42] The asymmetric broadening of $A_1$(LO) phonon mode in the InN is a prominent signature for the high carrier density. In the present case, $A_1$(LO) phonon mode, observed in the frequency range of 589-594 cm$^{-1}$ (Fig. 4a), shows that the samples grown at 580-600 °C have symmetric behaviour of the $A_1$(LO) mode. On the other hand, asymmetric broadening of the $A_1$(LO) phonon mode was found to start for samples grown at 620 °C and above. Hence, it is proved that strong coupling of carrier with the LO phonon is observed in samples grown above 620 °C because of the possible presence of high carrier density in these samples. Origin of the high carrier density is attributed to the presence large number of $V_N$, as a result of partial decomposition of InN at high temperatures and we have already discussed it in the previous section. The spectral region of $A_1$(LO) phonon mode is shown in the figure 6a, which illustrates the symmetric and asymmetric behaviour of the mode in the 580-600 °C and 620-650 °C, respectively. The film grown at 560 °C also shows the asymmetric behaviour of $A_1$(LO) mode which is similar that observed for the high temperature grown films. It may be because of the reason that films grown at 560 °C have the poor crystalline quality. In contrary to this, high temperature grown films have the high crystalline quality.

Usually strain in the films are calculated from the shift of $E_2$(high) phonon mode rather than that for the $A_1$(LO) mode because of strong influence of carrier concentration in the later. It has been reported that frequency of $A_1$(LO) mode shifts towards higher energy side (blue shift) with increasing carrier density.[36] Interestingly, we observed a red shift for the $A_1$(LO) phonon mode with increasing growth temperature even though carrier concentration was increased. Peak positions of $A_1$(LO) modes were found using Wire-3.4 (Renishaw) peak pick software. The $A_1$(LO) mode is red shifted by 4.2 cm$^{-1}$ from 580 to 650 °C (Fig. 6b). The value is greater than the shift (1.8 cm$^{-1}$) observed for the $E_2$(high) phonon mode (Fig. 5a). Thus, it can be inferred that the peak position of $A_1$(LO) mode is more influenced by relaxation of strain rather than that with the carrier density.

Information obtained from Raman spectroscopic analysis using a microscopic objective is always localized. Area of microanalysis depends on the wavelength of light and numerical aperture of the objective lens used. However, to get the complete information of strain in films, one has to carry the XRD studies. Figure 6c shows the XRD pattern of InN films grown at two extreme growth conditions such as 580 and 650 °C. Here we have shown 2θ values with three strong reflections of (10-10), (0002), (10-11) only. All diffraction peaks are well matched with the wurtzite InN phase (JCPDF card of 00-050-1239). However, in case of 580 °C grown sample, the 2θ positions are towards higher side as compared to the strain free values of InN. Moreover, these values are shifted towards lower side in case 650 °C grown samples as compared to the 580 °C grown sample. This observation further elucidated the fact that the compressive strain existed in the film gown at 580 °C. As the growth temperature was increased, the strain relaxation occurred leading to lowering of the 2θ values *i.e.*, increase in the *d* value. In the case of biaxial strain relaxation, (10–10) and (0002) reflection peak positions must show the opposite trend. However, a similar trend in the shift of the 2θ values is observed for all reflections (Fig. 6c) emphasising the fact that the strain relaxation takes place through a different mechanism of hydrostatic in nature. Hence, XRD study also strongly corroborates the observation made in Raman spectroscopic analysis. The XRD data for four significant samples including the peak corresponding to substrate is also shown in the supplementary information (Fig. S1).

### 3.4 Photoluminescence spectroscopic study

After a long debate, optical band gap of InN has been settled at 0.75 eV in the near IR region.[4] However, reports of optical emission from the CVD grown InN phase is scarce. Earlier, band gap observed in high energy range (1.9 eV) was related to the energy shift due to the BM effect as a consequence of high carrier density originated because of the poor crystalline quality of InN films. The energy shift due to the BM effect is attributed to the occupation of electrons much above the conduction band minimum from where optical transition takes place.[43-44] From the free electron theory, magnitude of the energy shift due to the BM effect is defined as,[45]

$$\Delta_{BM} = \frac{\hbar^2}{2m^*}(3\pi^2 n_e)^{2/3} \quad \text{...............................................} (5)$$

Figure 7a shows the PL spectra measured at 80K for InN films grown at different temperatures in the range of 580−650 °C. To the best our knowledge, for the first time, InN films grown using APCVD technique have shown optical emission which requires high crystalline quality. InN film grown at 560 °C has not shown any distinct emission other than mere background which attributes to its poor crystalline quality (not shown in figure). However, in the present study band edge emission is largely blue shifted as compared to the reported band gap value of 0.75 eV.[4] The PL peak is centred at 1.2 eV for the 580 °C grown sample (Fig. 7a). The observed blue shift of 0.45 eV, is attributed to the combined effect of compressive strain as well as the BM shift because of the considerably high carrier density in the films. As the growth temperature increases, carrier density is observed to increase which is revealed from asymmetric broadening of $A_1$(LO) mode (Fig. 6a). To find actual shift due to the BM effect, one has to discard the strain effects in the films. As discussed earlier, high temperature grown sample are most relaxed films and one can neglect the strain related shift in the band edge emission. In case 650 °C grown sample, the band edge peaked at 0.89 eV, so that shift in the bad gap due to the BM effect was 0.14 eV. Assuming negligible contribution of strain in the final blue shift of the PL emission for sample grown at 650 °C, approximate carrier density is calculated ~ 4.49×10$^{18}$ cm$^{-3}$ ($m^*$ = 0.07 $m_e$) using the Eqn. (5). In case of low temperature grown sample, the band edge is further shifted higher energy side even though these samples having the low carrier density. Blue shift in the band edge is a consequence of the compressive strain existed in films. Compressive strain in the films dominates over the BM shift in case of the low carrier density samples. Moreover, the FWHM of band edge emission was found to be decreased in case of low temperature grown samples as compared to the high temperature grown samples (Fig. 7a). The increase in the FWHM with increasing growth temperature is attributed to the increased amount of $V_N$ related defects which eventually contribute to the carrier density. Increase in carrier density eventually results in to the band filling effect which leads to broadening of the optical emission lines. From this observation we reiterate that residual strain also plays a crucial role in deciding the band gap of InN along with the shift due to the BM effect. In present study, compressive strain in the films further shifted the band edge emission along with the BM effect.

Figures 7b-d show the temperature dependent PL spectra of the samples grown at 580, 600, and 650 °C respectively. The PL spectra are fitted with Gaussian function. The 580 °C grown sample shows (Fig. 7b) two prominent peaks with the energies of 1.159 ($E_1$) and 1.169 eV ($E_2$). Interestingly, as the temperature increases $E_2$ peak is diminished even though it is stronger than $E_1$ at low temperatures. This observation suggests that $E_2$ peak may correspond to the free exciton of the wurtzite InN phase.[46,47] The $E_1$ peak may correspond to band edge emission of the InN. However, peak position of $E_1$ does not show any variation with temperature, indicating behaviour of the band edge peak as degenerate InN.[48] Similar kind of trend is also observed in case of 600 °C grown sample (Fig. 7c). However, 650 °C grown sample shows (Fig. 7d) similar PL emission along with the increased amount of integrated intensity. An additional peak in the low energy side is also observed, represented by $E_3$. As the temperature increases, the intensity of $E_3$ peak comes down and only the band edge emission is found to dominate at high temperatures. It suggests that $E_3$ peak is due to the presence of point defects and may be related to $V_N$, which acts as a shallow donor as in case of GaN.[49] Disappearance of $E_3$ peak at high temperature is attributed to the complete excitation of all carriers to conduction band from shallow donors. Presence of $V_N$ in the InN sample with high carrier concentration is already discussed in previous section. Interestingly, integrated PL intensity is found to reduce drastically for all the samples as the temperature increases (supplementary information, Fig. S2). However, rate of decrease in the integrated intensity is more in case of high temperature grown samples as compared to that for the low temperature grown samples. This may be attributed to the increase in Auger recombination in samples with high carrier density or as a result of surface electron accumulation.[50,51] In case of surface electron accumulation, localized holes move towards surface where they recombine non-radiatively with large number of electrons. This observation suggests that effect of surface electron accumulation is prominent in case high temperature grown samples.

**Conclusion**

In conclusion, for the first time optically high quality InN films are grown in the APCVD technique. Difficulties in synthesis of InN in APCVD are surmounted by approaching the novel method of self-seeded catalytic approach. Raman spectroscopic analysis of $E_2$(high) and $A_1$(LO) modes are carried for the study the strain and carrier density in these films. In the present study, $A_1$(LO) phonon mode is more influenced by the biaxial compressive strain rather than that for the $E_2$(high) phonon mode. Compressive strain, resulted from the lattice and thermal expansion mismatch, is relaxed at high temperature by hydrostatic tensile strain produced from the N vacancies. Strain relaxation at high temperature is further corroborated by the red shift in the optical emission line. The carrier density of ~ $4.49 \times 10^{18}$ cm$^{-3}$ is calculated using the Burstein-Moss energy shift considering the final blue shift of the 650 °C grown.

**Acknowledgement**

We thank A. K. Tyagi, SND, IGCAR for his encouragement and A. K. Sivadasan, SND, IGCAR for his help in the growth of InN films.

**Figure Captions:**

**Figure 1.** AFM topographical images of InN films grown at a) 560 °C, b) 580 °C, c) 600 °C and d) 650 °C showing the mosaic nature.

**Figure 2.** Growth of InN with two different ramp schemes with process steps inscribed in the insets a) Scheme 1: No deposition observed b) Scheme 2: Successful deposition of InN observed involving four different process steps.

**Figure 3.** AFM topography of the sapphire substrate a) prior to deposition and b) after the process B showing uniform nucleation of seed nanoparticles.

**Figure 4.** a) Raman spectra of InN films grown at different temperatures showing the symmetric and asymmetric nature of $A_1$(LO) mode along with other symmetry allowed and non-zone centre phonon modes b) $E_2$(high) phonon mode (open circles) spectral region with Lorentzian fit (solid red curve).

**Figure 5.** a) Change in the $E_2$(high) phonon mode frequency (left side scale) and its FWHM (right side scale) with respect to temperature. b) Calculated strain relaxation with the growth temperature.

**Figure 6.** a) The spectral region of $A_1$(LO) mode. Lines are guide to an eye. b) Dependence of $A_1$(LO) mode frequency with temperature, c) XRD pattern of samples grown at 580 and 650 °C temperatures (first three reflections shown here).

**Figure 7.** a) PL spectra, measured at 80K, for InN films grown at different temperatures. The spectra show the red shift of emission line with increase in temperature. b) - d) shows the temperature dependent PL spectra of 580, 600, and 650 °C grown samples with Gaussian fit for b) and d).

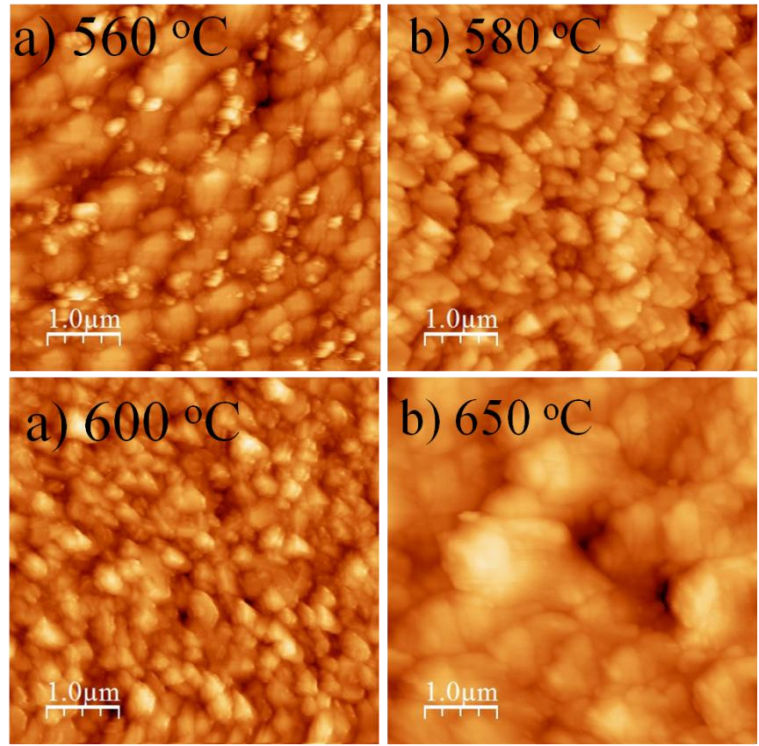

**Figure 1.** AFM topographical images of InN films grown at a) 560 °C, b) 580 °C, c) 600 °C and d) 650 °C showing the mosaic nature.

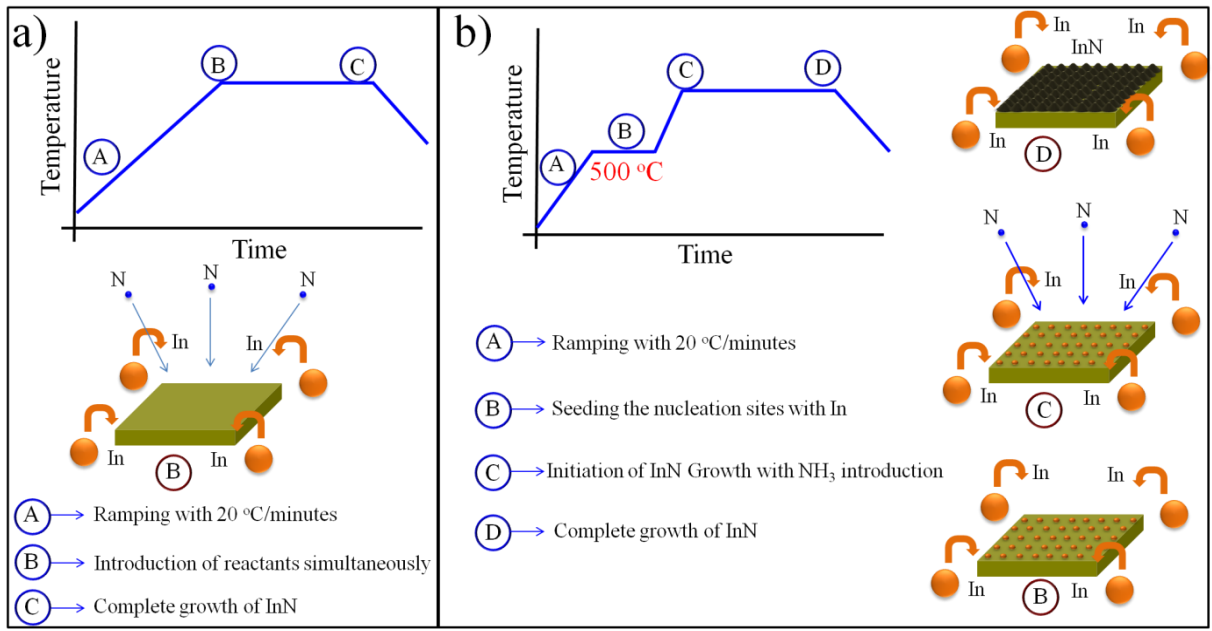

**Figure 2.** Growth of InN with two different ramp schemes with process steps inscribed in the insets a) Scheme 1: No deposition observed b) Scheme 2: Successful deposition of InN observed involving four different process steps.

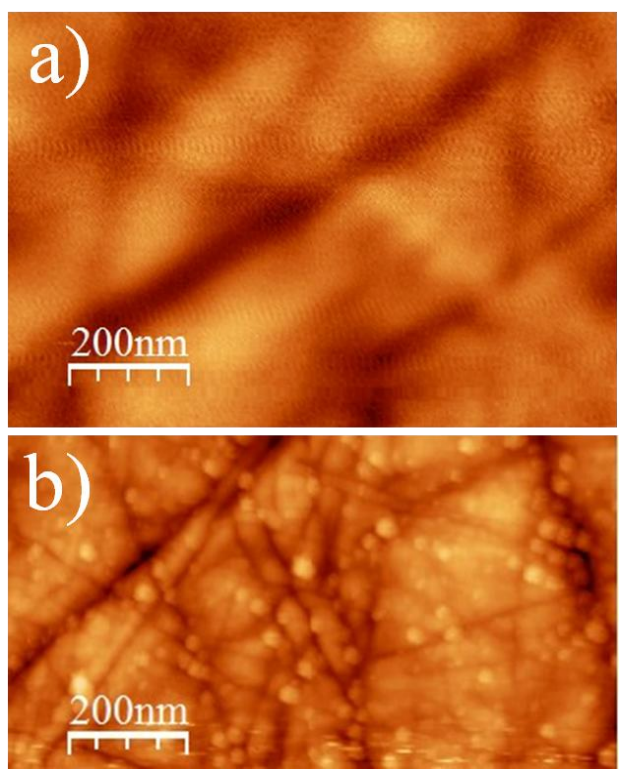

**Figure 3.** AFM topography of the sapphire substrate a) prior to deposition and b) after the process B showing uniform nucleation of seed nanoparticles.

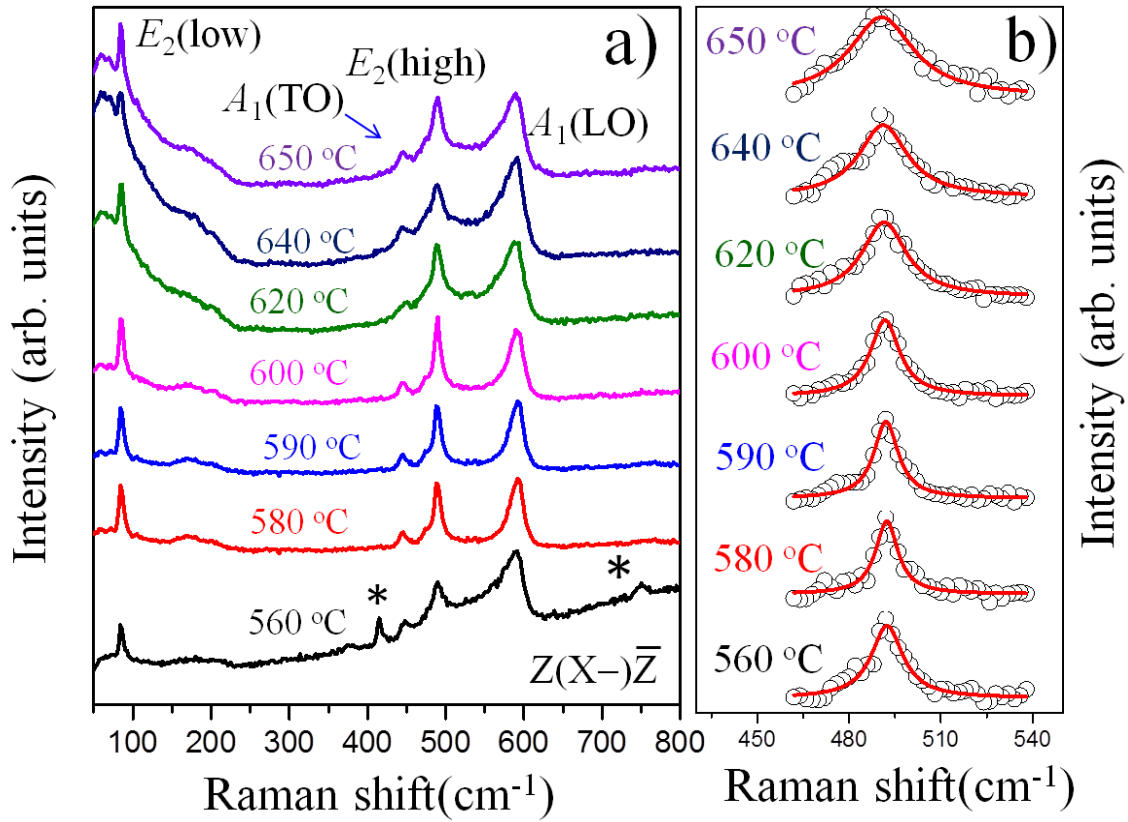

**Figure 4.** a) Raman spectra of InN films grown at different temperatures showing the symmetric and asymmetric nature of $A_1$(LO) mode along with other symmetry allowed and non-zone centre phonon modes b) $E_2$(high) phonon mode (open circles) spectral region with Lorentzian fit (solid red curve).

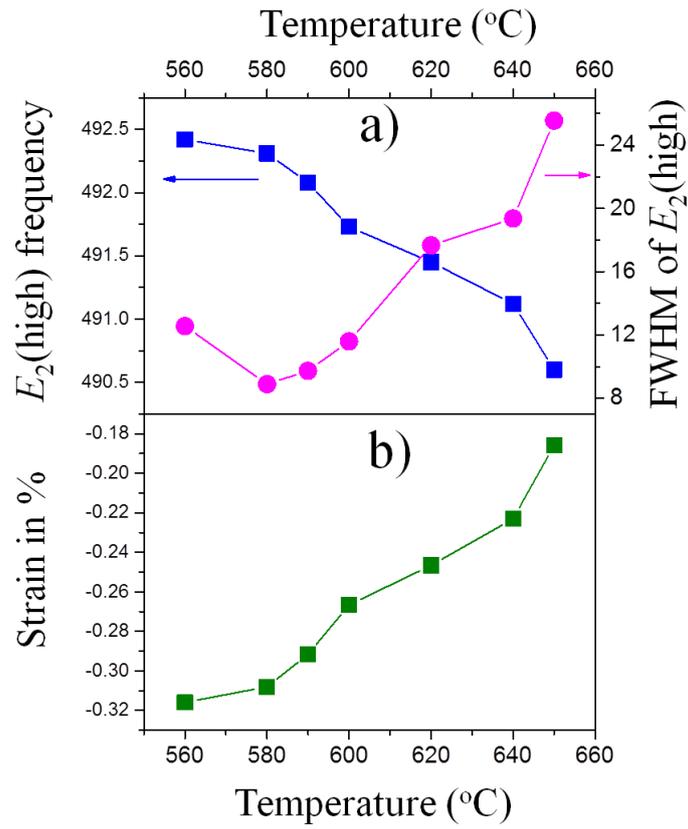

**Figure 5.** a) Change in the $E_2$(high) phonon mode frequency (left side scale) and its FWHM (right side scale) with respect to temperature. b) Calculated strain relaxation with the growth temperature.

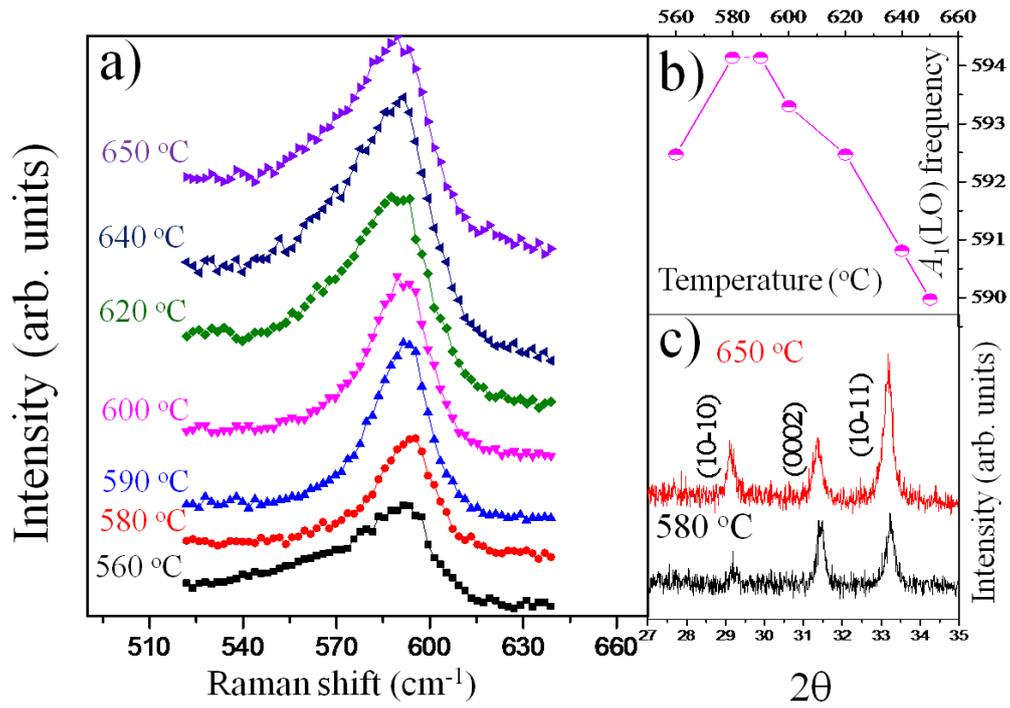

**Figure 6.** a) The spectral region of $A_1$(LO) mode. Lines are guide to an eye. b) Dependence of $A_1$(LO) mode frequency with temperature, c) XRD pattern of samples grown at 580 and 650 °C temperature (first three reflections shown here).

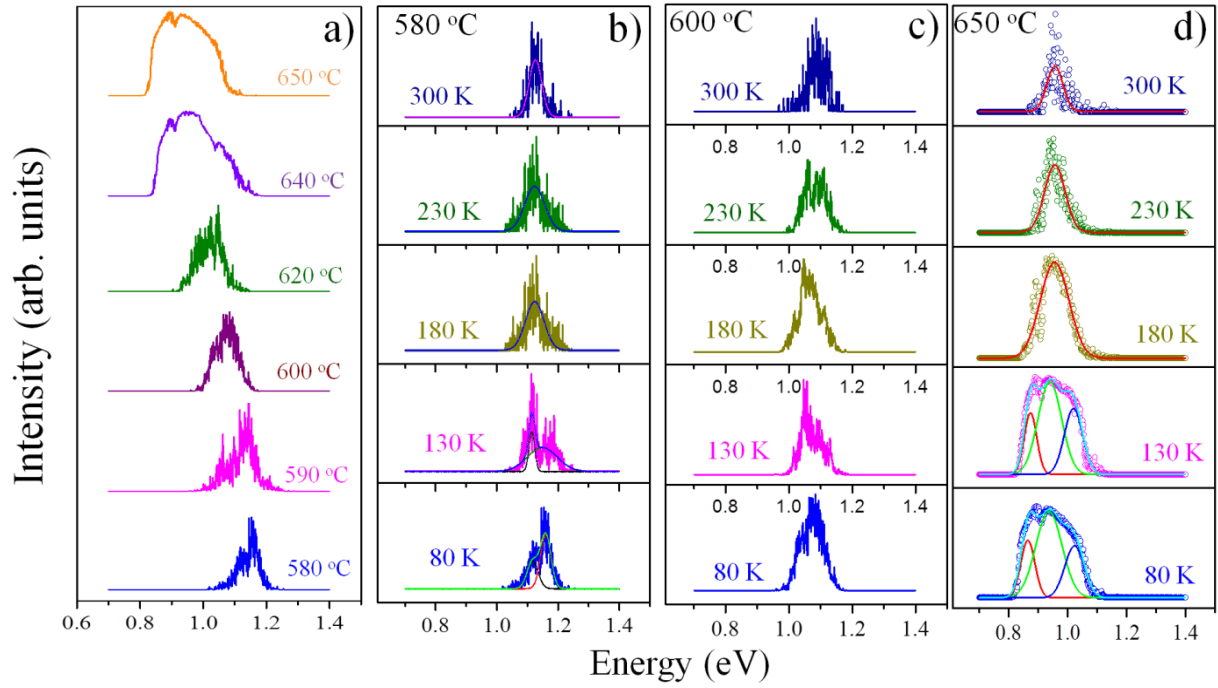

**Figure 7.** a) PL spectra, measured at 80K, for InN films grown at different temperatures. The spectra show the red shift of emission line with increase in temperature. b) - d) shows the temperature dependent PL spectra of 580, 600, and 650 °C grown samples with Gaussian fit for b) and d).

Supporting Information:

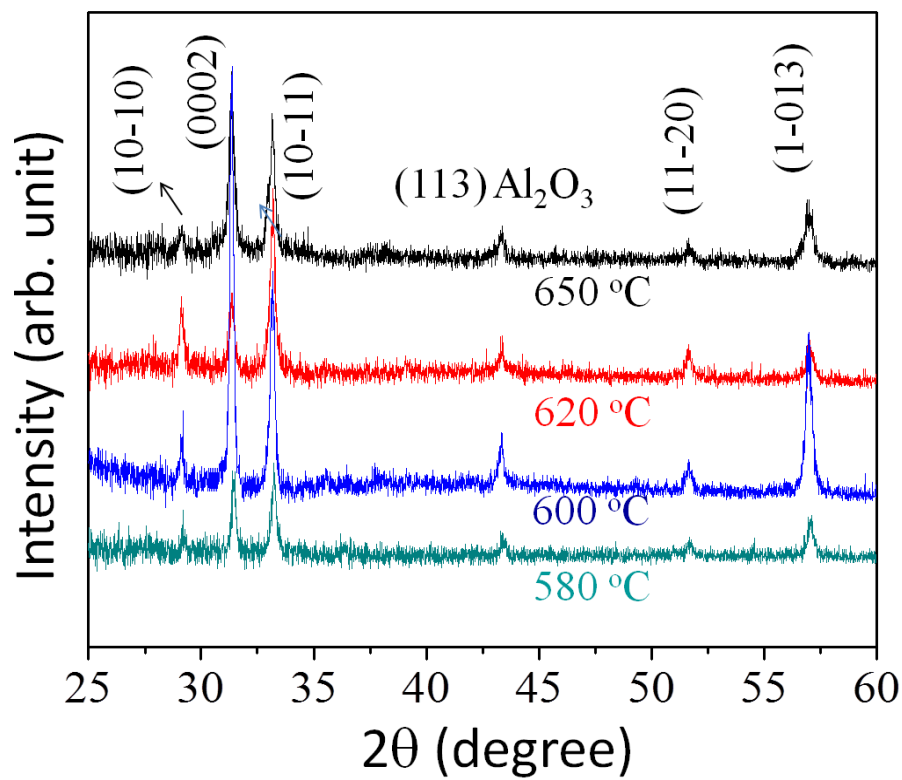

**Figure S1.** XRD pattern of InN samples grown at different temperatures.

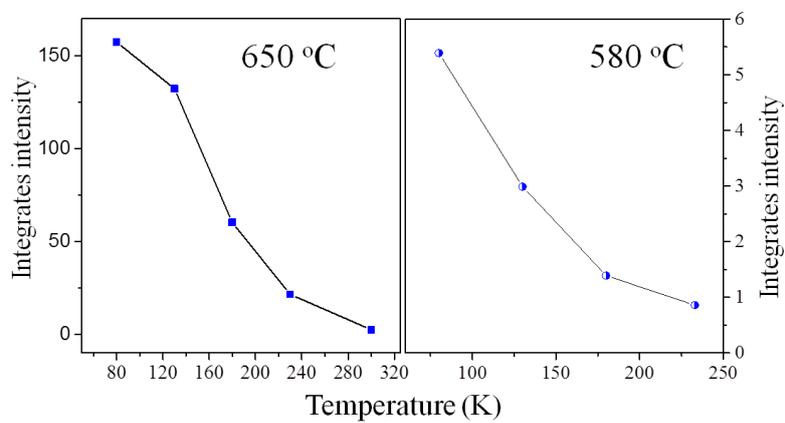

**Figure S2**. Temperature dependence of integrated PL intensity.